\documentstyle[aps,multicol]{revtex}
\begin{document}
\draft
\title{Exact effective action for fermions in one dimension with backscattering
at a boundary}
\author{Manuel Fuentes$^{1,2}$, Ana Lopez$^2$ and Eduardo Fradkin$^1$}
\address{$^1$Department of Physics, University of Illinois at
Urbana-\-Champaign , 1110 West Green Street,Urbana, IL 61801-3080 , USA.\\
$^2$ Department of Physics , Theoretical Physics , Oxford University , 1 Keble
Rd. ,
Oxford , OX1 3NP , United Kingdom.}

\maketitle

\begin{abstract}
We report exact results for  the partition function for free Dirac fermions
on a half line with physically sensible boundary conditions.
An exact effective action for
general backscattering amplitudes is derived. The  action also includes the
effects of both a (time-dependent) forward scattering amplitude and  a
dynamical
chiral twist of the fermion boundary conditions.
For a small backscattering amplitude, the effective
action has the expected boundary  Sine-Gordon form.
We discuss applications of our results to one-dimensional Fermi systems
with local backscattering.
\end{abstract}
\bigskip


\begin{multicols}{2}

\columnseprule 0pt

\narrowtext

%%%%%%%%%%%%%
\def\slp{{\raise.15ex\hbox{$/$}\kern-.57em\hbox{$\partial$}}}
%%%%%%%%%%%%%
%%%%%%%%%%%%%%%%%%%%%
\def\lnA{\raise.15ex\hbox{$/$}\kern-.57em\hbox{$A$}}
\def\slD{\raise.15ex\hbox{$/$}\kern-.57em\hbox{$D$}}
\def\slB{\raise.15ex\hbox{$/$}\kern-.57em\hbox{$B$}}
\def\sls{\raise.15ex\hbox{$/$}\kern-.57em\hbox{$s$}}
\def\slbarA{{\raise.15ex\hbox{$/$}\kern-.57em\hbox{$\bar A$}}}
\def\sla{\raise.15ex\hbox{$/$}\kern-.57em\hbox{$a$}}
\def\slb{\raise.15ex\hbox{$/$}\kern-.57em\hbox{$b$}}
%%%%%%%%%%%%%%%%%%%%

In the past few years there has been considerable interest in the study of
one-dimensional Fermi systems coupled to quantum impurities.
The (by now) standard approach to this problem was  formulated by Kane
and Fisher~\cite{Kane}. It consists of first finding a bosonized
form of the theory (with suitably chosen boundary conditions), to account for
Luttinger liquid effects, and then to consider the
effects of scattering with the (quantum) impurity in the already bosonized
theory.
By means of standard bosonization methods it is a simple matter to study the
effects of forward scattering processes at the impurity. However, both
abelian~\cite{llm} and non-abelian bosonization~\cite{witten}
can deal with backscattering processes ({\it i.~e.\/}, local mass terms)
only in a perturbation expansion in the mass or backscattering
terms~\cite{cn}.
Recently, we have shown~\cite{paper1} that it is possible to use path-integral
methods~\cite{laplata} to attack problems with Dirac fermions in
systems with boundaries~\cite{falomir} coupled to boundary degrees of freedom.
In this paper we consider the effects of backscattering terms in
one-dimensional Fermi systems without the use of mass perturbation theory. We
make
extensive use of the well known equivalence~\cite{emery} between
non-relativistic
fermions at finite density ( with Fermi momentum $k_F$) and the theory of
massless
Dirac fermions in $1+1$ dimensions (with Fermi velocity $v_F=1$). For
simplicity we
consider only free spinless fermions although the methods can be extended
easily to
include the effects of Luttinger-type interactions. The quantum impurity
is represented
by a boundary  {\it forward scattering} amplitude $a_0$ ( which we will regard
as
the time
component of a vector potential at the boundary), a boundary mass-like
{\it backscattering}
potential $M$  and a {\it chiral angle} $\theta$ (which modifies the fermion
boundary
conditions),
all with an arbitrary time dependence. For physically reasonable systems,
the forward and backward scattering amplitudes $a_0$ and $M$ have the same time
dependence
(if any) since they represent the $Q\approx 0$ and $Q\approx 2 k_F$ components
of the
scattering amplitudes at the {\it same} quantum impurity.
The chiral angle $\theta$ is an external degree of freedom which modifies the
fermion
boundary conditions by changing the amplitudes of the left and right movers at
the
boundary by a time-dependent relative phase. Thus, all physical observables
must be
periodic in this angle. Its main effect is to generate local fluctuations of
the fermion
density near the boundary.

The main result of this paper is the derivation of an {\it exact} effective
action
of the system of fermions, for the boundary degrees of freedom $M$, $a_0$ and
$\theta$.
We follow the methods outlined in reference~\cite{paper1}, where we
applied  a technique introduced by Forman~\cite{forman}
for the computation of fermion determinants with boundaries.
We discuss the form
of the effective action in several interesting regimes.
The full
non-linear, {\it exact} effective action
reduces to well known expressions in the limit of small mass.
We also show that the exact
effective action is, as required, periodic in the chiral angle.
We see that,
when backscattering is present, the
chiral angle and the forward scattering amplitude are not equivalent
descriptions of
the quantum impurity.

Throughout this paper we work in two-dimensional euclidean space.
Due to technicalities associated with hermiticity properties of the euclidean
Dirac
operator, in addition to the usual analytic continuation to imaginary time
$x_0 \to -i x_2$, we adopt the standard prescription~\cite{laplata} which also
continues into the complex plane the chiral angle
$\theta \to i \theta$ and the time component of the vector potential $a_0 \to i
s_2$.
This procedure has the unwelcome feature that the group of chiral
transformations
become non-compact in euclidean space ({\it i.~e.\/} the chiral group has been
complexified). Thus, at the end of the calculation we will continue  back
$\theta$ in
order to recover physically transparent results.
For the sake of generality we also include a spacial component $s_1$ which will
turn
out to play no role.
The (euclidean) Lagrangian for the system is
\begin{equation}
{\cal L}_{F}={\bar \psi} \;i  \slp \; \psi +{\bar \psi}\; \gamma_\mu \psi \;
A^{\rm imp}_\mu(x)
+i  M(x) \; {\bar \psi} \psi
\label{eq:Lth}
\end{equation}
where $  A^{\rm imp}_\mu(x)$, $M(x)$ represent the
scattering of the electrons by the impurity. The first ({\it vector})
scattering term
represents the forward
scattering amplitude (for the underlying lattice or non-relativistic system)
and it
does not mix Left and  Right movers.  $A^{\rm imp}_\mu$ has the form
$A^{\rm imp}_\mu(x)= \delta(x_1) \; s_\mu(x_2)$
since the systems of physical interest have a (possibly
time-dependent) degree of freedom localized at $x_1=0$.
The  remaining amplitude represents the backscattering processes (which mix
Left  with
Right) .
We take as the boundary condition for the fermi field $R(0,x_2)=-
e^{2\theta(x_2)}L(0,x_2)$\cite{paper1}. Notice that we have already continued
$\theta$.
The effective action is obtained from the functional integral as follows
\begin{eqnarray}
e^{-S_{\rm eff}(\theta,s_\mu,M)}&=& \int {\cal D} {\bar \psi} {\cal D} \psi
\exp \left(-\int d^2x {\cal L}_F \right) \nonumber \\
&\equiv&{\rm Det}{(  i \slp+ \sls + iM)}
\label{eq:functint}
\end{eqnarray}
Therefore, we have to calculate  the determinant of the
$(1+{1\over{2}})$-dimensional Dirac operator coupled to a $\delta(x_1)$
massive potential and to a gauge field $s_\mu(x_2)\delta(x_1)$ .
Following  Forman's method \cite{paper1,forman}, we reformulate the
determinant of the operator we
want to compute, as the determinant of a different operator but with
suitable chosen boundary conditions.
The formal way to compute the determinant of this operator is by computing
its eigenvalues. In order to do so it is useful to  regularize the $\delta$
function as
the limit $\epsilon \rightarrow 0$ of the function $v_{\epsilon}(x)$ where
$\int_{\epsilon - {\delta \over 2}}^ {\epsilon + {\delta \over 2}}
v_{\epsilon}(x)=1$.
It is clear that when $\epsilon,\delta \rightarrow 0$, the solutions of
$[ i \slp + \sls +iM]\psi=\lambda \psi$
in $x_1\geq0$ will be the same as the solutions of the equation
$i \slp \psi=\lambda \psi$
plus a suitable chosen boundary condition.
Integrating the first equation over the interval $(\epsilon - {\delta \over 2},
\epsilon + {\delta \over 2})$ and assuming that the field $\psi$ is finite
in such interval, we obtain for small $\delta$
\begin{equation}
\psi({\epsilon}^{+})= e^{{\gamma}_5s_2(x_2) +i s_1(x_2)-{\gamma}_1M(x_2)}
\psi({\epsilon}^{-}).
\end{equation}
Recall that at $\epsilon^-$ the fields satisfy $Be^{-
\gamma_5\theta(x_2)}\psi(\epsilon^-)=0$ where $B$ is the $2 \times 2$ matrix
$B=I+\sigma_1$.
In the limit $\epsilon \rightarrow0$, the boundary condition for the
field $\psi$ becomes
$B{\cal U}^{-1}(x_2)\psi|_{x_1=0}=0$,
where ${\cal U}(x_2)= e^{[{\gamma}_5s_2(x_2) -
{\gamma}_1M(x_2)]}e^{{\gamma}_5\theta(x_2)}$.
Thus, we have to compute the determinant of the Dirac operator with twisted
boundary
conditions.

In order to apply Forman's methods we introduce an auxiliar parameter $\tau$
such that
${\cal U}(\tau)= e^{\tau[{\gamma}_5s_2(x_2) -
{\gamma}_1M(x_2)]}e^{\tau{\gamma}_5\theta(x_2)}$.
We must consider a complete, but
otherwise arbitrary, system of functions in the kernel of this differential
operator, which do not need to satisfy any particular boundary condition.
As in \cite{paper1} we take the basis
\begin{eqnarray}
\psi_{n}(x_2,x_1)=e^ {iw_{n}[x_2-i\gamma_{5}(x_1-L)]}\left(
\begin{array}{cc}
1 \\
1
\end{array} \right)\;
,\; n\in
{\mathchoice {\hbox{$\sf\textstyle Z\kern-0.4em Z$}}
{\hbox{$\sf\textstyle Z\kern-0.4em Z$}}
{\hbox{$\sf\scriptstyle Z\kern-0.3em Z$}}
{\hbox{$\sf\scriptscriptstyle Z\kern-0.2em Z$}}}
\end{eqnarray}
where $w_{n}= {{(2n+1)\pi}\over {T}}$.
We now define the new basis vectors
$h_{\mu}^{n} (x_2)\equiv B {\cal U} ^{-1} (\mu) \psi_{n}(x_2)$
where $\psi_n(x_2)= \psi_n(0,x_2)$.
We expand the vectors $h_{\mu}^{n} (x_2)$ and $h_{\mu + \tau}^{n} (x_2)$
in  a complete set of functions in $[0,L]$ which we choose to be the basis
$ \psi_{n}(L,x_2)=  \exp( iw_{n} x_2)$, with
$w_{n}= {{(2n+1)\pi}\over {T}}$
$( n\in
{\mathchoice {\hbox{$\sf\textstyle Z\kern-0.4em Z$}}
{\hbox{$\sf\textstyle Z\kern-0.4em Z$}}
{\hbox{$\sf\scriptstyle Z\kern-0.3em Z$}}
{\hbox{$\sf\scriptscriptstyle Z\kern-0.2em Z$}}}
)$.
The $m$-th component of $h_{\tau}^{n} (x_2)$ is
\begin{equation}
h_{\tau}^{n,m} (x_2) = {1\over {2T}} \int ^{T\over {2}}_{-T\over {2}}  dx_2 \;
e^ {[ i(w_{n}-w_{m}) x_2]} \;h_{\mu}^{n} (x_2)
\end{equation}

Forman's Theorem \cite{forman} states that there exists a matrix
$\Phi_{\mu}(\tau)$
(to be specified below) such that
\begin{equation}
{d\over {d\mu}}\ln\;\left({
{\rm Det} ( i \slp )_{B{\cal U}^{-1}(\mu + \tau)}}
\over {{\rm Det} ( i \slp)_{B{\cal U}^{-1}(\mu)}}
\right) =
{d\over {d\mu}}\ln\; \det \Phi_{\mu}(\tau).
\label{eq:forman}
\end{equation}
The determinants on the left-hand side are defined through the $\zeta$
-function regularization, while the right-hand side is a well-defined quantity.
We define the operator $\Phi_{\mu}(\tau)\equiv h(\mu +\tau) h^{-1}(\tau)$
satisfying
\begin{eqnarray}
{d\over {d\mu}} \ln\; \det \Phi_{\mu}(\tau)=
{\rm Tr} &&\left[h^{-1}(\mu+\tau){d\over {d\mu}} h( \mu+\tau)\right.\nonumber\\
&& -\left.
h^{-1}(\mu){d\over {d\mu}}h( \mu) \right]
\end{eqnarray}
To compute $h(\tau)$ we recall that $h_{\tau}^{n}= B{\cal
U}^{-1}(\tau)\psi_{n}$, therefore it has
an expansion in positive and negative frequency modes.In particular, in the
limit $L\rightarrow \infty$ it becomes ( up to a normalization constant)
\begin{equation}
h_{\tau}^{n} (x_2)=H_{+}(\tau)\Theta(w_n)+H_{-}(\tau)\Theta(-w_n)
\end{equation}
where $\Theta(w)$ is the step function, and we have defined
$H_{\pm}(\tau)$ to be
\begin{equation}
H_{\pm}=\!
(\cosh(\tau\Phi) \pm {s_2\over {\Phi}}\sinh(\tau\Phi))
e^{\pm \tau \theta}
\mp i{M\over {\Phi}}\sinh(\tau\Phi)e^{\mp\tau \theta}
\label{eq:Hpm}
\end{equation}
with $\Phi(\tau,x_2)= \sqrt{ M^2(x_2) + s_2^2(x_2)}$.
Thus, $h(\tau)$ is a matrix of the form
\begin{equation}
h(\mu) \: =
\: \left(
\begin{array}{cc}
{[e^{-\phi_1}]}_{--} & {[e^{\phi_2}]}_{-+} \\
{[e^{-\phi_1}]}_{+-} & {[e^{\phi_2}]}_{++}
\end{array} \;\; \right)\;
\end{equation}
where
\begin{eqnarray}
\phi_1(\tau,x_2)&=&-\ln H_{-}(\tau)\nonumber \\
\phi_2(\tau,x_2)&=&+\ln H_{+}(\tau)
\label{eq:phi1}
\end{eqnarray}
Following the same algebra described in\cite{paper1} we get
\begin{eqnarray}
{\rm Tr}[{d\over{d\tau}}&&h(\tau) h^{-1}(\tau)] =
{\rm tr} \left\{ -[{d\over{d\tau}}\phi_1]_{++} +[{d\over{d\tau}}\phi_2]_{--}
\right.
\nonumber\\
&&+\left. [{d\over{d\tau}}\phi_1]_{+-}({[e^{(\phi_1+\phi_2)_-}]}_{-+}
{[e^{-(\phi_1+\phi_2)_-}]}_{++} )\right.
\nonumber\\
&&+\left.
[{d\over{d\tau}}\phi_2]_{-+}({[e^{-(\phi_1+\phi_2)_+}]}_{++}
{[e^{(\phi_1+\phi_2)_+}]}_{+-})
\right\}
\label{eq:traza}
\end{eqnarray}
In the last expression, $(\phi_1+\phi_2)_-$ denotes the part of
$(\phi_1+\phi_2)$ with negative frequencies, and $(\phi_1+\phi_2)_+$
the one with positive
frequencies. The first two terms yield in the $T\rightarrow \infty$ limit
\begin{equation}
H_0(\tau)
= -{1\over {4\pi a}}\int dx_2
{\partial_{\tau}}(\phi_1(x_2) -\phi_2(x_2))
\label{eq:H_0}
\end{equation}
where $a$ is a short-distance regulator.

The last two terms yield, after anti-transforming Fourier, the following
functional
\begin{equation}
H_1(\tau)\!= {1 \over {4\pi}}\int \!dx\!\int\! dy
{\partial_{\tau}}(\phi_1+\phi_2) (x)
K(x,y) (\phi_1+\phi_2)(y)
\label{eq:H_1}
\end{equation}
where $K(x,y)= -{1\over \pi} {\cal P} {1\over {(x - y)^2}} +
{1\over a} \delta (x -y)$.
 Hence
\begin{eqnarray}
{d\over {d\mu}}ln\!&&\!\left({
{\rm Det} ( i \slp )_{B{\cal U}^{-1}(\mu + \tau)}}
\over {{\rm Det} ( i \slp)_{B{\cal U}^{-1}(\mu)}}
\right) =\nonumber\\
&=&H_0(\mu + \tau) + H_1(\mu +\tau)-H_0(\mu) - H_1(\mu)
\end{eqnarray}
Following reference~\cite{falomir}, we find that
\begin{equation}
{\rm Det}{( i \slp)}_{B{\cal U}^{-1}}=\exp \int^1_0\;d\tau
(H_0(\tau) + H_1(\tau))
\label{eq:det1}
\end{equation}
Substituting Eqs.~(\ref{eq:H_0}) and (\ref{eq:H_1}) into Eq.~(\ref{eq:det1}),
and
using that $H_0(0)=H_1(0)=0$, we obtain
\begin{eqnarray}
&&\ln{\rm Det}{( i \slp)}_{B{\cal U}^{-1}} =
 -{1\over {4\pi a}}\int dx (\phi_1(x) -\phi_2(x)) \nonumber\\
&+&{1 \over {8\pi}}\int dx\int dy (\phi_1+\phi_2)(x)
K(x,y) (\phi_1+\phi_2)(y)
\end{eqnarray}
Therefore, the effective action, Eq.(\ref{eq:functint}), becomes
\begin{eqnarray}
&S&_{\rm eff}(\theta,s_2,M)={-1\over {4\pi a}}\int \ln\left(
\cosh^2 \Phi
 + \frac{M^2-s_2^2}{\Phi^2}\sinh^2 \Phi \right. \nonumber\\
&+& \left. {\frac{2iM}{\Phi}}
\sinh \Phi \; [\cosh \Phi \; \sinh(2 \theta ) + {s_2 \over \Phi}
\sinh \Phi \; \cosh(2 \theta )]\right)\nonumber\\
&-&{1 \over {8\pi}}\int dx\int dy \; U(x) \; K(x,y)\; U(y)
\label{eq:effac}
\end{eqnarray}
where $U(x)$ is defined by the expression
\begin{equation}
U \equiv
\ln \frac{(\cosh \Phi +{s_2 \over
{\Phi}}\sinh \Phi )e^{ \theta}- i{M \over {\Phi}} \sinh \Phi \;
e^{-\theta})}{(\cosh \Phi -
{s_2 \over {\Phi}}\sinh \Phi)e^{- \theta}+ i{M \over {\Phi}} \sinh \Phi
\; e^{\theta})}
\label{eq:U}
\end{equation}
Eq.~(\ref{eq:effac}) is the {\it exact}  effective action for a
system of Euclidean Dirac fermions coupled to a boundary backscattering
potential.
This action is completely written in terms of the boundary degrees of
freedom, {\it i.~e.\/}, in terms of $\theta(x_2)$, $s_2(x_2)$, and
$M(x_2)$. We will
discuss now several limiting regimes and the analytic continuation back to
Minkowski
space.

In order to check the consistency of our result, we begin by
comparing the
expression for the effective action of Eq.(\ref{eq:effac}) in the
small $M$ limit , with the effective action obtained using standard
perturbation theory in the mass $M$. For simplicity we consider the case where
the
forward scattering amplitude  is absent, $s_2(x_2)=0$. After setting
$s_2(x_2)=0$ in
the definitions of $\phi_1$ and $\phi_2$ in eq.~(\ref{eq:phi1}) ,
it is easy to see that, to leading order in $M$, the effective action of
eq.~(\ref{eq:effac}) becomes
\begin{eqnarray}
&&S_{\rm eff}(\theta,0,M)=
-{1\over {2\pi}}\int dx \int dy\; \theta (x) K(x,y)\; \theta (y)
\nonumber\\
&& +{i\over {\pi}}\int dx \int dy\; M(x)\cosh(2\theta(x))
\;K(x,y)\; \theta(y)
\nonumber\\
&&-{i\over {2\pi a}}\int dx \;M(x) \sinh(2\theta(x))
\label{eq:masachica}
\end{eqnarray}
We stress here that this limit has been obtained directly from the exact result
{}.
Now, we calculate the expansion of the fermionic determinant in
Eq.(\ref{eq:functint}) in
powers of $M$. Using that, for any operator ${\cal O}$, $\ln {\rm
{\rm Det}{\cal O}} = {\rm Tr} \ln {\cal O}$, the effective action becomes
\begin{eqnarray}
S_{\rm eff}(\theta,s_2,M&)&=-{\rm Tr} \ln {( i \slp+ \sls + iM)}
\nonumber\\
=&-&{\rm Tr} \ln \left[ ( i \slp+ \sls)
(1 + S_F(x,y|\theta,s_\mu) M)\right]
\end{eqnarray}
where  $S_F(x,y|\theta,s_\mu)$  is the fermionic Green's function which
satisfies
\begin{equation}
 \left\{
\begin{array}{ll}
(i\slp_x  + \sls)S_F(x,y|\theta,s_\mu)= \delta(x-y)\\
S^{11}_F(0,x_2;y|\theta,s_\mu)=- e^{2\theta(x_2)}S^{21}_F(0,x_2;y|\theta,s_\mu)
{}.
\end{array}
\right.
\label{eq:Stheta}
\end{equation}
Hence, after expanding the logarithm in powers of $M$ and keeping only the
leading order term, we obtain
\begin{equation}
e^{-S_{\rm eff}(\theta,s_\mu)}={\rm Det}{( i \slp
+ \sls)} e^{{\rm Tr}\;(iMS_F)}
\label{eq:saprox}
\end{equation}
{}From now on we will set $s_\mu\equiv0$. To compute the exponential in the
above
equation we replace $S_F(x,y|\theta)$ by the one particle Green's
function for free fermions on the half line calculated in
reference \cite{paper1}. Then
\begin{eqnarray}
{\rm Tr}(iMS_F)={i\over {2\pi}}&&\!\int \!d^2x  M(x)  \left(
\frac{e^{{1\over {\pi}}\int dz_2 \theta(z_2)h_{11}(z_2)}}{(x_1+y_1) + i(x_2-
y_2)} \right. \nonumber \\
&-&\!\left. \left. \frac{e^{{1\over {\pi}}\int dz_2
\theta(z_2)h_{22}(z_2)}}{(x_1+y_1) -
i(x_2- y_2)}\right)\!\right|_{\lim x\rightarrow y}
\end{eqnarray}
where $h_{11}$ and $h_{22}$ were defined in reference \cite{paper1}.
After some lengthy, but otherwise straightforward algebra, it can be
shown that, in the
particular case of $M(x)=\lim_{a\rightarrow 0} M(x_2)\delta(x_1-a)$, the above
equation
becomes \cite{foot1}
\begin{eqnarray}
&&{\rm Tr}\;(iMS_F)= {\frac{i}{2\pi a}}\int
dx\;M(x)\sinh(2\theta(x))\nonumber\\
&&-{\frac{i}{\pi}}\int dx \int dy \;M(x) \cosh(2\theta(x))K(x,y)
\theta (y)
\label{eq:even}
\end{eqnarray}
The final form for the effective action in the small mass limit is now easily
found
by recalling that the first factor in Eq.(\ref{eq:saprox}) is the fermion
determinant calculated in reference \cite{paper1},
\begin{equation}
{\rm Det}{( i \slp)} =\exp\left({1\over {2\pi}}\int
dx \int dy\;\theta(x)\;K(x,y)\;\theta (y)\right)
\label{eq:det2}
\end{equation}
After replacing Eqs.~(\ref{eq:even}) and (\ref{eq:det2}) into
Eq.~(\ref{eq:saprox}),
we recover  the expression for the  effective action of
Eq.~(\ref{eq:masachica}).

Next we consider the analytic continuation of the boundary angle $\theta$.
Since in
the exact effective action the chiral angle always enters in the form of an
exponential $\exp(\pm \theta)$, (in practice only $\exp(2 \theta)$ ever
appears),
the analytic continuation $\theta \to i \theta$ effectively recompactifies the
chiral
group thus making the physics {\it periodic} in $\theta$ ( with period $\pi$).
By undoing the analytic continuation $\theta \to i \theta$ (while remaining
in imaginary time), and after some algebra, Eq.~(\ref{eq:masachica}) becomes
\begin{eqnarray}
&&S_{\rm eff}(\theta,0,M)={1\over {2\pi}}\int dx \int dy\;
\theta(x) \;K(x,y)\; \theta(y)
\nonumber\\
&+&{1\over {2\pi a}}\int dx \;M(x)
\sin\left[2\theta(x)-2M(x)\cos(2\theta(x))\right]
\label{eq:masachica3}
\end{eqnarray}
where we replaced $\theta \to \theta|_{{\rm mod} \; \pi}$.  Thus, the effective
action is
explicitly periodic in $\theta$. Notice that there is no periodicity in $a_0$
or in $M$.
By keeping only the lowest orders in $M$ we find
\begin{eqnarray}
&&S_{\rm eff}(\theta,0,M)\approx {1\over {2\pi}}\int dx \int dy\;
\theta(x) \;K(x,y)\; \theta(y)
\nonumber\\
&+&\!{1\over {2\pi a}}\!\int \!dx\!\left[\!M(x) \sin(2\theta(x))
+M^2(x)\cos^2(2\theta(x))\! \right]
\label{eq:masachica4}
\end{eqnarray}
Up to a shift of $\theta \to \theta+ {\frac{\pi}{4}}$, we recognize that first
two
terms in eq.~(\ref{eq:masachica4}) are the Kane-Fisher bosonized form
of the
action~\cite{Kane} to lowest order in the mass $M$. The correction terms are
higher
harmonics in $\theta$ which presumably can be neglected since (at least
perturbatively in $M$) they are irrelevant in an Renormalization Group sense.
Notice that $M$ may actually vary with time, as it would in an X-ray edge
process.
This action has also been studied extensively in the context of the
Caldeira-Leggett
problem~\cite{caldeira}. An exact solution of this system, based on the
Thermodynamic
Bethe Anstaz, has been given recently~\cite{fendley}.

Finally, we discuss the regime $\theta=s_2=0$ and $M(x_2)$ arbitrary.
In this regime we find the effective action for $M$ to be given by
\begin{eqnarray}
&&S_{\rm eff}(M)=-{1\over {4\pi a}}\int dx\; \ln \cosh(2 M(x))
\nonumber\\
&+&\!\!{1\over {2\pi}}\!\!\!\int \!\!\!dx \!\!\!\int\!\!\!dy\tan^{-1}\!\tanh
M\!(x)
K(x,y) \tan^{-1}\!\tanh M\!(\!y)\!\!
\label{eq:masasola}
\end{eqnarray}
Notice that the first term breaks the aparent periodicity in $M$ which, unlike
the
chiral degree of freedom, should not be present. The scale of this term is
set by the parameter $a$ which is the short distance cutoff (the only scale
left!).
It is interesting to notice that the limits $M \to 0$ and $s_2 \to 0$ do not
commute.

In summary, in this paper we derived an exact, non-linear, non-perturbative
effective
action for a quantum impurity coupled to a Fermi field in
$1+{1\over 2}$-dimensions. The  effective action is a non-linear function  of
the
backscattering amplitude $M$ and forward scattering amplitude $a_0$ of the
fermions
and of a chiral angle $\theta$ which twists the fermion boundary conditions.
The effective
action is explicitly periodic in  $\theta$ but its not  in $a_0$ or in $M$.
Physically this is
a very natural result. The chiral angle shifts the relative phase of right and
left movers
and, as such, the physics has to be periodic in $\theta$. On the other hand,
both  $M$
and $a_0$ are local scattering amplitudes that enter in the Hamiltonian. Their
effects
include local density fluctuations of the Fermi system and non-trivial
contributions to
phase shifts. Hence, the physics is not periodic in these amplitudes.
In the limit of small $M$ we recover the boundary Sine-Gordon interaction.
Elsewhere we report on generalizations of  our results to include the effects
of fermion
correlations\cite{elsewhere}.

This work was supported in part
by the National Science Foundation through the grants NSF DMR94-24511 at
the University of Illinois at Urbana-Champaign,
NSF DMR-89-20538/24 at the Materials Research Laboratory of the University of
Illinois(MF,EF), by a Glasstone Research Fellowship in Sciences and a Wolfson
Junior
Research Fellowship (AL).

%\newpage

% Begin PRL format
\end{multicols}
% End PRL format


\begin{references}
\bibitem{foot1} Here we used the prescription
$\lim_{x_1\rightarrow 0} \frac{e^{-|w|x_1}}{x_1}=\lim_{a\rightarrow 0}
({1\over a}-|w|)$.
This is consistent with the regularization  used in reference \cite{paper1}.
\end{references}
\end{document}